\documentclass[onecolumn,showpacs,amsmath,amssymb,prd,nofootinbib]{revtex4}
\def\sss{\scriptscriptstyle}
\def\^#1{^{\sss #1}}
\def\_#1{_{\sss #1}}
\def\beq{\begin{equation}}
\def\eeqno#1{\label{#1}\end{equation}}
\def\ten#1#2{^{\sss#1}_{\sss#2}}

\def\az{a\_{0}}

\def\l0{\ell\_{0}}

\def\rar{\rightarrow}
\def\s{\sigma}
\def\l{\lambda}
\def\o{\omega}

\def\f{\phi}

\def\t{\theta}
\def\k{\kappa}
\def\z{\zeta}

\def\r{\rho}

\def\m{\mu}
\def\n{\nu}
\def\Up{\Upsilon}
\def\C{\Gamma}

\def\L{\mathcal{L}}

\def\M{\mathcal{M}}

\def\T{\mathcal{T}}
\def\Tmn{\T\_{\m\n}}
\def\Th{\hat{\mathcal{T}}}
\def\hTmn{\Th\_{\m\n}}

\def\Sh{\hat{\mathcal{S}}}
\def\Sb{\bar{\mathcal{S}}}
\def\St{\tilde{\mathcal{S}}}

\def\D{\Delta}

\def\d{\delta}

\def\a{\alpha}
\def\b{\beta}
\def\c{\gamma}
\def\d{\delta}
\def\eps{\epsilon}
\def\vr{{\bf r}}
\def\vx{{\bf x}}

\def\vv{{\bf v}}

\def\vn{{\bf n}}

\def\grad{\vec\nabla}
\def\div{\vec \nabla\cdot}
\def\gf{\grad\phi}
\def\fpg{4\pi G}

\def\emn{\eta\_{\m\n}}
\def\Emn{\eta\^{\m\n}}
\def\gmn{g\_{\m\n}}
\def\Gmn{g\^{\mu \nu}}

\def\Gab{g\^{\alpha\beta}}

\def\Glr{g\^{\lambda\rho}}

\def\hgh{\hat g^{1/2}}
\def\gh{g^{1/2}}

\def\fh{\hat\f}
\def\gfh{\grad\fh}
\def\fs{\f^*}
\def\gfs{\grad\fs}
\def\hs{h^*}

\def\hC{\hat\C}
\def\cd#1{{}_{\sss;#1}}
\def\mcd#1{{}_{\sss:#1}}

\def\km{\k\_-}
\def\kp{\k\_+}
\def\oo{\lambda}
\def\sl{(}
\def\sr{)}
\begin{document}

\title{Bimetric MOND gravity}
\author{Mordehai Milgrom }
\affiliation{ The Weizmann Institute Center for Astrophysics}
%\address{ The Weizmann Institute Center for Astrophysics}
\begin{abstract}
A new relativistic formulation of MOND is advanced,
involving two metrics as independent degrees of freedom: the MOND
metric $\gmn$, to which alone matter couples, and an auxiliary
metric $\hat g\_{\m\n}$. The main idea hinges on the fact that we
can form tensors from the difference of the Levi-Civita connections
of the two metrics,
$C\ten{\a}{\b\c}=\C\ten{\a}{\b\c}-\hC\ten{\a}{\b\c}$, and these act
like gravitational accelerations. In the context of MOND  we can
form dimensionless `acceleration' scalars, and functions thereof
(containing only first derivatives) from contractions of
$\az^{-1}C\ten{\a}{\b\c}$. I look at a class of bimetric
MOND theories governed by the action  $I=-(16\pi G)^{-1}\int[\b\gh R
+ \a\hgh \hat R
 -2(g\hat g)^{1/4}f(\k)\az^2\M(\tilde\Up/\az^2)]d^4x
+I\_M(\gmn,\psi_i)+\hat I\_M(\hat g\_{\m\n},\chi_i),$ with
$\tilde\Up$ a scalar quadratic in the $C\ten{\a}{\b\c}$, $\k=(g/\hat
g)^{1/4}$, $I\_M$ the matter action, and allowing for the existence
of twin matter that couples to $\hat g\_{\m\n}$ alone. Thus, gravity
is modified not by modifying the `elasticity' of the space-time in
which matter lives, but by the interaction between that space-time
and the auxiliary one. In particular, I concentrate on the
interesting and simple choice $\tilde\Up\propto
\Gmn(C\ten{\c}{\m\l}C\ten{\l}{\n\c}
-C\ten{\c}{\m\n}C\ten{\l}{\l\c})$. This theory introduces only one
new constant, $\az$; it tends simply to general relativity (GR) in
the limit $\az\rar 0$, and to a phenomenologically valid MOND theory
in the nonrelativistic limit. The theory naturally gives MOND and
``dark energy'' effects from the same term in the action, both
controlled by the MOND constant $\az$. As regards gravitational
lensing by nonrelativistic systems--a holy grail for relativistic
MOND theories--the theory predicts that the same potential that
controls massive-particle motion also dictates lensing in the same
way as in GR: Lensing and massive-particle probing of galactic
fields will require the same ``halo'' of dark matter to explain the
departure of the present theory from GR. This last result can be
modified with other choices of $\tilde\Up$, but lensing is still
enhanced and MOND-like, with an effective logarithmic potential.
\end{abstract}
%\keywords{}
\maketitle

\section{\label{introduction}Introduction}
From the inception of MOND \cite{milgrom83} it has been clear that
the paradigm needs buttressing by a relativistic formulation.
Indeed, efforts to construct such a formulation started shortly
thereafter, with the tensor-scalar version sketched in \cite{bm84}.
This was the first in a chain of theories of increasing force,
culminating in the advent of the tensor-vector-scalar theory (TeVeS)
of Bekenstein \cite{bek04}. Some landmarks along this track are
described in \cite{sanders97,bek04,bek06,z06,z07,skordis09}; see, in
particular, the reviews in \cite{bek06,skordis09}. All these
theories involve as independent degrees of freedom an Einstein
metric, whose free action is the standard Einstein-Hilbert action,
with additional scalar and/or vector degrees of freedom, with their
own actions. These scalar/vector degrees of freedom are used to
dress up the Einstein metric into the `physical' metric to which
matter couples. TeVeS has a version of the nonrelativistic (NR)
theory proposed in \cite{bm84} as a NR limit.
\par
Another line of relativistic theories that aim to reproduce MOND
phenomenology has been propounded in \cite{blt08,blt09}, based on
the omnipresence of a gravitationally polarizable medium proposed in
\cite{bl07}.
\par
Here I propound a new class of relativistic formulations for the
MOND paradigm in the form of bimetric MOND (BIMOND) theories. These
came to light as follows: I have recently described
\cite{milgrom09c} a new class of nonrelativistic, bi-potential MOND
theories, a subclass of which is governed by a Lagrangian density of
the form
  \beq \L=-{1\over 8\pi
G}\{\b(\gf)^2+\a(\gfh)^2-\az^2\M[(\gf-\gfh)^2/\az^2]\}
  +\r({1\over 2}\vv^2-\f),  \eeqno{futcol}
  leading to the field  equations
  $$\div[\m^*(|\gfs|/\az)\gfs]=\fpg\r,
  ~~~~~~~  \m^*(y)\equiv\b-{\a+\b\over\a}\M'(y^2)$$
    \beq \D\f=\div[(1-\a^{-1}\M')\gf^*]
  =\fpg\b^{-1}\r+\b^{-1}\div(\M'\gf^*), \eeqno{hutred}
with $\fs=\f-\fh$. I also described in detail the requirements from
$\a,~\b$, and $\M(z)$ that lead to the required MOND and Newtonian
limits of these theories. In particular, I discussed at length the
interesting case $\b+\a=0$ ($\b=1$ then normalizes $G$ to be the
Newton constant), which leads to the field equations
  \beq \D\f^*=\fpg\r,
~~~~~~\D\f=\div[(1+\M')\gf^*]=\fpg\r+\div(\M'\gf^*), \eeqno{hutram}
with $\M'$ a function of $(\gfs/\az)^2$, such that $\M'(z)\rar 0$
for $z\rar \infty$ ensures the Newtonian limit, and $\M'(z)\approx
z^{-1/4}$ in the MOND regime $z\ll 1$. This is a particularly
tractable MOND theory, as it requires solving only linear
differential equations, with the inevitable MOND nonlinearity
entering only algebraically. In all the NR theories above, matter
couples only to one of the potentials: the MOND potential $\f$,
while $\fh$ is an auxiliary potential, and in the special case of
Eq.(\ref{hutram}) their difference $\fs$ is exactly the Newtonian
potential of the problem.
\par
These NR MOND theories have inspired the construction of closely
analogous relativistic MOND theories with two metrics as
independent, gravitational degrees of freedom, which I begin to
investigate here.\footnote{In these theories the two metrics are
independent degrees of freedom. Theories like Brans-Dicke, TeVeS,
etc., are also sometimes described as being bimetric because they
involve two metrics, but those two metrics are a priori related
conformally or disformally via other degrees of freedom such as
scalars or vectors.} This new class of BIMOND theories involve only
$\az$ as a new constant. They tend to general relativity (GR) in the
limit $\az\rar 0$, which is a desirable trait. And, they tend to a
MOND theory compatible with MOND phenomenology in their NR limit.
\par
These theories, like all other relativistic versions of MOND
proposed to date, must, I believe, be only approximate, effective
theories to be derived from some more fundamental picture that
underlies them. This is pointed to by the appearance of an a priori
unspecified function in all these theories.
\par
The use of two (or more) metrics to describe gravity has a long
history. For example, Rosen \cite{rosen74} considered bimetric
theories, where the auxiliary metric is forced to be flat. More
recently, it was found \cite{boulanger01}  that ghosts appear in a
large class of bimetric theories (apparently not including the
present BIMOND). More matter-of-principle questions regarding
bimetric gravities are discussed in
\cite{dk02,bdg06,bdg07,banados09}, but these authors
 confined themselves to metric couplings that
involve only the metrics, not their derivatives, as in the case of
BIMOND.
\par
In section \ref{formalism}, I present the formalism underlying the
BIMOND theories; in section \ref{nonrelativistic}, I consider the NR
limit of these theories, showing how they lead to NR MOND theories;
section \ref{gr} demonstrates how the theories go to GR in the limit
$\az\rar 0$; section \ref{lensing} discusses lensing; section
\ref{cosmology} discusses cosmology briefly, and section
\ref{discussion} is a discussion.

\section{\label{formalism}Formalism}
 The NR theories mentioned above involve two
potentials, the MOND potential $\f$ felt by matter, and an auxiliary
one $\fh$. They point to relativistic BIMOND theories involving the
MOND metric $\gmn$, to which matter couples, and which in the NR
limit reduces to $\f$, and an auxiliary metric $\hat g\_{\m\n}$.
\par
Working with two metrics enables us to form nontrivial tensors and
scalars from the difference in their Levi-Civita connections
$\C\ten{\a}{\b\c}$ and $\hC\ten{\a}{\b\c}$,
 \beq C\ten{\a}{\b\c}=\C\ten{\a}{\b\c}-\hC\ten{\a}{\b\c},
  \eeqno{veyo}
involving only first derivatives of the metrics, which is not
possible with a single metric. This is particularly pertinent in the
context of MOND, since connections act like gravitational
accelerations. So, without introducing new constants in the
relativistic formulation we can write Lagrangian functions of
dimensionless scalars constructed from $\az^{-1}C\ten{\a}{\b\c}$
that enable us to interpolate between the GR limit, $\az\rar 0$, and
the MOND limit, $\az\rar \infty$.
\par
The tensor $C\ten{\a}{\b\c}$ is related to covariant derivatives of
one metric with the connection of the other (more generally, they
relate covariant derivatives of tensors with respect to the two
connections):
 \beq \gmn\mcd{\l}=g\_{\a\n}
C\ten{\a}{\m\l}+g\_{\a\m}C\ten{\a}{\n\l}~~~~~~ \hat
g\_{\m\n}\cd{\l}=-\hat g\_{\a\n} C\ten{\a}{\m\l} -\hat
g\_{\a\m}C\ten{\a}{\n\l},\eeqno{berda}
 \beq C\ten{\l}{\a\b}={1\over 2}\Glr(g\_{\a\r}\mcd{\b}
+g\_{\b\r}\mcd{\a} -g\_{\a\b}\mcd{\r}) =-{1\over 2}\hat
g\^{\l\r}(\hat g\_{\a\r}\cd{\b} +\hat g\_{\b\r}\cd{\a} -\hat
g\_{\a\b}\cd{\r}),
 \eeqno{tured}
where the covariant derivative $(;)$ is taken with the connection
$\C\ten{\a}{\b\c}$ and $(:)$ with $\hat\C\ten{\a}{\b\c}$. We can
form various scalars out of $C\ten{\a}{\b\c}$ and the metrics. One
scalar that will be of particular use to us is based on the tensor
 \beq \Up\_{\m\n}=C\ten{\c}{\m\l}C\ten{\l}{\n\c}
 -C\ten{\c}{\m\n}C\ten{\l}{\l\c},  \eeqno{kurta}
with the same index combination that appears in the expression for
the Ricci tensor
 \beq R\_{\m\n}=\C\ten{\a}{\m\a,\n}-\C\ten{\a}{\m\n,\a}+
\C\ten{\c}{\m\l}\C\ten{\l}{\n\c}
 -\C\ten{\c}{\m\n}\C\ten{\l}{\l\c},  \eeqno{ricci}
 and in $\hat R\_{\m\n}$ constructed similarly from $\hat g\_{\m\n}$.
 One finds
 \beq R\_{\m\n}-\hat R\_{\m\n}=C\ten{\l}{\m\l;\n}
-C\ten{\l}{\m\n;\l}-\Up\_{\m\n}.\eeqno{nista} Thus, using well known
manipulations, the scalar $\Up\equiv\Gmn\Up\_{\m\n}$ connects the
two Ricci scalars $R=\Gmn R\_{\m\n}$ and the mixed $\hat R\_m=\Gmn
\hat R\_{\m\n}$ by
 \beq R-\hat R\_m=-\Up+g^{-1/2}(\gh\Gmn
 C\ten{\l}{\m\l})\_{,\n}
-g^{-1/2}(\gh\Gmn C\ten{\l}{\m\n})\_{,\l}. \eeqno{nerol}  Similarly,
interchanging the roles of $\gmn$ and $\hat g\_{\m\n}$,
 \beq \hat R- R\_m=-\hat\Up-\hat g^{-1/2}(\hgh\hat g\^{\m\n}
 C\ten{\l}{\m\l})\_{,\n}
+\hat g^{-1/2}(\hgh\hat g\^{\m\n} C\ten{\l}{\m\n})\_{,\l},
\eeqno{nerolos} where $\hat\Up=\hat g\^{\m\n}\Up\_{\m\n}$,
$R\_m=\hat g\^{\m\n} R\_{\m\n}$, $\hat R$ is the Ricci scalar of
$\hat g\_{\m\n}$, and $g$ and $\hat g$ are minus the determinants of
$\gmn$ and $\hat g\_{\m\n}$ respectively.
\par
We can construct gravitational Lagrangian densities using the
scalars $R,~\hat R,~R\_m,~\hat R\_m$, and scalars constructed by
contracting powers of $C\ten{\l}{\m\n}$ with the two metrics and
their inverses (there are also $\hat g/g$, $\bar\o\equiv\Gmn\hat
g\_{\m\n}$, etc. that can be used). If we only contract with $\gmn$
and $\Gmn$, a quadratic scalar is a linear combination (possibly
with coefficients depending on scalars such as $g/\hat g$ or
$\bar\o$) of the following scalars
 \beq \Gmn C\ten{\c}{\m\l}C\ten{\l}{\n\c},~~~\bar C\^{\c}C\_{\c},~~~
 \gmn\bar C\^{\m}\bar C\^{\n},~~~\Gmn C\_{\m} C\_{\n},
 ~~~g\_{\a\l}g\^{\b\m} g\^{\c\n}C\ten{\a}{\b\c}C\ten{\l}{\m\n},
 \eeqno{meqaz}
 where
$\bar C\^{\c}\equiv \Gmn C\ten{\c}{\m\n}$, $C\_{\c}\equiv
C\ten{\a}{\c\a}$. The choice of scalars to be used may be forced on
us by various theoretical and phenomenological desiderata (see
below). The main point is that $C\ten{\c}{\m\n}$ have only first
derivatives of the metrics, that they reduce to derivatives of the
potential difference in the Newtonian limit (in the sense to be
discussed below), and that we can form dimensionless quantities from
them with the MOND acceleration $\az$ (or a MOND length
$\ell=c^2/\az$). As regards the four curvature scalars, is will be
advantageous to include them in the action only linearly, and eschew
terms such as in the voguish $f(R)$ theories. Such nonlinear terms
render the theory a higher derivative one, which I would like to
avoid.\footnote{For the same reason I avoid scalars that are higher
order in the curvature tensors, such as the different possible
contractions of $R\_{\m\n}$ with itself or with $\hat R\_{\m\n}$.
These are even less appealing as explained in \cite{woodard07}.}
Another reason to avoid such terms in the MOND context is that they
do not naturally lead in their NR limit to a single constant $\az$
controlling the dynamics.\footnote{For example, to account for
dimensions correctly, a function of $R$ has to be introduced as
$f(\ell^2 R)$, with $\ell$ some length scale. The NR limit of $R$
includes $c^{-2}\D\f$ as the dominant term in $\f/c^2$, and second
order ones such as $c^{-4}(\gf)^2$, and $c^{-4}\f\D\f$. Thus, in the
argument of $f$ the second term will give $(\gf/\az)^2$, with
$\az=c^2/\ell$, which fits well into the MOND frame. But the first,
dominant, term would involve a time scale $\ell/c$, not an
acceleration. When $R$ appears linearly, the first term becomes
immaterial in the action, as a complete derivative, and we are left
with terms that are welcome in MOND [the $\f\D\f$ term is also
$(\gf)^2$ up to a derivative].} Neither obstacle appears if we allow
functions of scalars made of $C\ten{\a}{\b\c}$. These contain only
first derivatives of the metrics, and give NR limits in which only
$\az$ appears (see below). We see from Eqs.(\ref{nerol}) that $\gh
R$ and $\gh\hat R\_m$ differ by $\Up$ plus a total derivative so it
is enough to include one of these in the action, as we anyhow permit
functions of $\Up$. The same is true of the pair $\hgh R\_m$ and
$\hgh \hat R$. Because the number of possible combinations it too
large to explore here, I limit myself to the subclass of actions of
the form
 \beq I=-{c^4 \over 16\pi G}\int[\b \gh R +\a\hgh \hat R
 -2(g\hat g)^{1/4}f(\k)\ell^{-2}\M(\ell^m\Up\_{i}^{(m)})]d^4x
 +I\_M(\gmn,\psi_i)+\hat I\_M(\hat g\_{\m\n},\chi_i),  \eeqno{gedat}
where $\ell\equiv c^2/\az$, $\k\equiv(g/\hat g)^{1/4}$, $f(1)=1$,
and $\Up\_{i}^{(m)}$ are scalars formed by contracting a product of
(even) $m$ $C\ten{\a}{\b\c}$, which can be used in principle. In
what follows, I shall confine myself to quadratic
scalars.\footnote{The MOND constant $\az$ is normalized so that the
mass-asymptotic-velocity relation is $MG\az=V^4$. It defines the
scale length $\ell$ that is used in the coefficient and the argument
of $\M$. Any dimensionless factors can be absorbed in the definition
of $\M$ so that its coefficient is $\ell^{-2}=c^{-4}\az^2$ and its
argument is as prescribed here.} I have included two matter actions:
The first, $I\_M$, involves the matter degrees of freedom with which
we interact directly, designated symbolically as $\psi_i$. It
contains only the MOND metric $\gmn$ to which matter is coupled in
the standard way. The other, $\hat I\_M$, involves other matter
degrees of freedom, $\chi_i$, and only $\hat g\_{\m\n}$, to account
for the possibility that $\hat g\_{\m\n}$ controls a matter world of
its own. There are no direct (electromagnetic, etc.) interactions
between the $\psi$ matter and the twin $\chi$ matter.\footnote{To
obviate possible confusion, note that the twin matter is not to play
the role of the putative dark matter in galactic systems; this is
still fully replaced by MOND effects; see below.}
\par
I make two requirements of the action: a. Require that it gives a NR
MOND theory in its NR limit. This means the following: given a non
relativistic system of slow masses one can express the metrics
solution of the relativistic theory in terms of potentials so that
the equations of motion for slow particles in the resulting (multi)
potential theory are those  required by NR MOND, with the
appropriate MOND and Newtonian limits; this is a phenomenological
requirement (by itself it does not dictate the effects on massless
particles--e.g., gravitational lensing--even in  NR systems). b.
Require that the action gives GR in the limit $\az\rar 0$. This is
not a phenomenological necessary (for example, TeVeS does not
satisfy it), but I feel that it is highly desirable for various
reasons. This automatically causes the theory to agree with the
stringent constraints from the solar system and binary
pulsars--which are known to agree with GR--because the accelerations
in these systems are many orders of magnitude larger than $\az$. I
also require this limit lest we have to introduce additional
constant(s) to the theory, which has to give GR in some limit of its
parameters.
\par
When the two metrics are conformally related, which might be the
case in certain circumstances, $\gmn=e\^{\vartheta(x)}\hat
g\_{\m\n}$, we have $C\_{\l}=2\vartheta\_{,\l}$, $\bar
C\^{\l}=-\Glr\vartheta\_{,\r}$,
$\Up\_{\m\n}=(1/2)(\gmn\Gab\vartheta\_{,\a}\vartheta\_{,\b}
-\vartheta\_{,\m}\vartheta\_{,\n})$,
$\Up=(3/2)\Gmn\vartheta\_{,\m}\vartheta\_{,\n}$. If we a priori
constrain our metrics to be conformally related (i.e. vary the
action only over such pairs) we get the Brans-Dicke theory with the
choice $\M(z)\propto z$ (and appropriate choice of the constants and
$f(\k)$, and possibly using $\hat\Up$ in the argument of $\M$). With
a more general form of $\M(z)$, we then get the relativistic MOND
theory sketched in \cite{bm84}.
\par
Without the interaction $\M$ term, the theory separates into two
disjoint copies of GR. It is important to note that as a combined
structure, the theory then enjoys a larger symmetry involving
separate coordinate transformations in the two separate actions.
This double symmetry has to be brought to bear when solving the
field equations of the theory, which now satisfy two sets of Bianchi
identities. So, eight gauge conditions can, and have to, be
employed. It is the interaction that breaks this larger symmetry,
as, generically, it is only invariant to application of the same
coordinate transformation to the two metrics. However, under certain
circumstances the interaction is symmetric under a more extended set
of coordinate transformation, and we must be careful then to employ
the larger gauge freedom. The above mentioned complete decoupling is
an example that, as we shall see in section \ref{gr}, applies in the
formal limit $\az\rar 0$ of the theory (leading, as we want, to GR).
It may also happen, in principle, that the interaction term vanishes
only in some limited regions of space-time; for example, if the
extreme GR limit applies in some regions. In this case we must allow
for gauge freedom involving coordinate transformations that coincide
only outside these regions, but not inside them. We shall see
another example in section \ref{nonrelativistic}, where the NR limit
of the theory has such a partial double gauge freedom.

\subsection{Concrete simple example}
I shall hereafter concentrate on a simple special case of the class.
Some generalizations will be mentioned briefly below, in this
section, and in section \ref{cosmology}.
\par
In the first place, I take $\M$ to be a function of only one scalar,
quadratic in the $C\ten{\a}{\b\c}$. In particular, I find the scalar
$\Up$ defined above a natural choice for this argument, as it has
the same structure as the first-derivative part of the Ricci
curvature scalar (not itself a scalar)
 \beq \C\^{(2)}\equiv\Gmn(\C\ten{\c}{\m\l}\C\ten{\l}{\n\c}
 -\C\ten{\c}{\m\n}\C\ten{\l}{\l\c}).  \eeqno{vuted}
It is well known that one can replace $R$ in the Einstein-Hilbert
action by $\C\^{(2)}$ and still get GR. Here we can also do this,
replacing also $\hat R$ by the corresponding $\hat \C\^{(2)}$, and
making $\M$ a function of $\Up$, which is constructed in the same
way from $C\ten{\a}{\b\c}$. We shall also see that with this choice
of scalar argument the NR limit of the theory is especially simple.
\par
As a further simplification I take $\a+\b=0$. This will yield a
particularly interesting and simple subclass of theories, which turn
out to have the theory (\ref{hutram}) as their NR limit for slowly
moving masses in a double Minkowski background. I then take $\b=1$
for $G$ to be Newton's constant.
\par
Work in units in which $c=1$, and use $\az=\ell^{-1}$ to highlight
the connection with MOND. Also, anticipating the expression for NR
limit of $\Up$, I take the argument of $\M$ to be $-\Up/2\az^2$. The
relativistic action I then consider is
  \beq I=-{1\over 16\pi G}\int[\gh R
  -\hgh\hat R
-2(g\hat g)^{1/4}f(\k)\az^2\M(-\Up/2\az^2)]d^4x
 +I\_M(\gmn,\psi_i)-\hat I\_M(\hat g\_{\m\n},\chi_i).  \eeqno{gedap}
 [Using Eq.(\ref{nerolos}) we can replace the first two terms
by $(\gh\Gmn -\hgh\hat g\^{\m\n})R\_{\m\n}+\hgh\hat\Up$.] I take a
mixed volume element for the interaction term, with $f$ normalized
such that $f(1)=1$. Note the change of sign in the definition of the
twin matter action to match the negative sign for the
Hilbert-Einstein action of $\hat g\_{\m\n}$.
 \par
 Varying over $\Gmn$ and over $\hat g\^{\m\n}$ we get, respectively
 \beq  G\_{\m\n}+S\_{\m\n}=-8\pi G \Tmn,  \eeqno{misha}
 \beq \hat G\_{\m\n}+\hat S\_{\m\n}=-8\pi G \hTmn, \eeqno{nuvec}
where $G\_{\m\n}$ and $\hat G\_{\m\n}$ are the Einstein tensors of
the two metrics,
 \beq G\_{\m\n}=R\_{\m\n}-{1\over 2}R\gmn ,
 ~~~~~~~~\hat G\_{\m\n}=\hat R\_{\m\n}-{1\over 2}
  \hat R\hat g\_{\m\n},
\eeqno {einas} $\Tmn$ and $\hTmn$ are the matter energy-momentum
tensors (EMT); e.g.,  $\d I\_M\equiv-(1/2)\int\gh \Tmn\d\Gmn$, and
$S\_{\m\n},~\hat S\_{\m\n}$ are the functional derivatives (one with
an opposite sign) of the interaction term with respect to the two
metrics:
 \beq \d\int
-2(g\hat g)^{1/4}f(\k)\az^2\M(-\Up/2\az^2)d^4x \equiv \int(
\gh\d\Gmn S\_{\m\n}-\hgh\d\hat g\^{\m\n} \hat
S\_{\m\n})d^4x.\eeqno{gutrast} For the present choice of the scalar
argument of $\M$, we have
 \beq
 S\_{\m\n}=\km\M'\Up\_{\m\n}
 +(\km\M'\St\ten{\l}{\m\n})\cd{\l}-
 \Lambda\_m\gmn, \eeqno{cayul}
  \beq
 \hat S\_{\m\n}=(\kp\M'\Sh\ten{\l}{\m\n})\mcd{\l}-
 \hat\Lambda\_m\hat g\_{\m\n}, \eeqno{budra}
 \beq \St\ten{\l}{\m\n}=C\ten{\l}{\m\n}-\d\ten{\l}{\sl\m}C\_{\n\sr}
 +{1\over 2}(C\^{\l}-\bar C\^{\l})\gmn, \eeqno{gyrtaz}
  \beq\Sh\ten{\l}{\m\n}=q\ten{\a}{\sl\m}C\ten{\l}{\n\sr\a} +\Glr
C\ten{\a}{\r\sl\m}\hat g\_{\n\sr\a} -\hat\Glr
C\ten{\a}{\r\b}q\ten{\b}{\sl\m}\hat g\_{\n\sr\a}
-q\ten{\l}{\sl\m}C\_{\n\sr}+{1\over 2}\gmn^*\hat g\^{\l\a}
C\_{\a}-{1\over 2}\bar C\^{\l}\hat g\_{\m\n},\eeqno{tujas}
 \beq   \Lambda\_m=-{1\over 2\k}[\k f(\k)]'\az^2\M,~~~~~~~~
 \hat\Lambda\_m=-{\k^3\over 2}[\k^{-1} f(\k)]'\az^2\M. \eeqno{cypul}
 Here,
 \beq
 C\^{\l}\equiv g\^{\a\l}C\_{\a},~~~~
 \k\_{\pm}\equiv \k\^{\pm 1}f(\k),~~~~
 q\ten{\m}{\a}=\Gmn\hat g\_{\n\a},~~~~\gmn^*=\hat g\_{\a\m}\Gab\hat
g\_{\b\n}, \eeqno{dunet} and $(\m...\n)=\{\m...\n+\n...\m\}/2$
signifies symmetrization over the two indices,
\par
The tensor
 \beq \tilde T\_{\m\n}={1\over 8\pi G}[\km\M'\Up\_{\m\n}
 +(\km\M'\St\ten{\l}{\m\n})\cd{\l}]
  \eeqno{vytip}
may be viewed as the EMT of the phantom dark matter (DM); whereas
the $\Lambda\_m$ term may roughly be viewed as ``dark energy''. Note
that the last term in $\St\ten{\l}{\m\n}$, which contributes
${1\over 2}[\km\M'(C\^{\l}-\bar C\^{\l})]\cd{\l}\gmn$ may also
contribute to the dark energy due to its form. Define in analogy
with $\tilde T\_{\m\n}$
 \beq \hat T\_{\m\n}={1\over
8\pi G}(\kp\M'\Sh\ten{\l}{\m\n})\mcd{\l} .\eeqno{nuval}
\par
The Einstein tensors satisfy the usual Bianchi identities
$G\^\n\_{\m;\n}=\hat G\^\n\_{\m:\n}=0$,\footnote{For each tensor
indices are raise with the corresponding metric; so, e.g.,
$G\^\n\_\m=g\^{\n\a}G\_{\m\a},~\hat G\^\n\_\m=\hat g\^{\n\a}\hat
G\_{\m\a}$.} derivable from the invariance of the Einstein-Hilbert
actions to coordinate transformations. In addition, we have here,
for the general action (\ref{gedat}), a set of four identities
following from the fact that the mixed term is a scalar; these read
 \beq S\^\n\_{\m;\n}
 -\k^{-2}\hat S\^\n\_{\m:\n}=0.
   \eeqno{gytad}
\par
Given that the matter EMTs are divergence free (for matter degrees
of freedom satisfying their own equations of motion):
$\T\^\n\_{\m;\n}=\Th\^\n\_{\m:\n}=0$, the above identities imply
four differential identities satisfied by our 20 field equations. If
we write these equations as $Q\_{\m\n}=0$ and $\hat Q\_{\m\n}=0$,
respectively, then the four relations
 \beq Q\^\n\_{\m;\n}
 -\k^{-2}\hat Q\^\n\_{\m:\n}=0
  \eeqno{gymash}
hold identically, and, as usual, deprive us of four equations to
account for the fact that the solution can be determined only up to
a coordinate transformation. This seems to leave us with a tractable
Cauchy problem, although this require more careful
checking.\footnote{As a result of identities (\ref{gytad}) and the
Bianchi identities, the four expressions $G\^0\_\m +S\^0\_\m
-\k^{-2}(\hat G\^0\_\m +\hat S\^0\_\m)$ contain only up to first
time derivatives of the metric, and cannot be used to propagate the
problem in time. Instead, the initial conditions have to satisfy the
four equations $Q\^0\_\m-\k^{-2}\hat Q\^0\_\m=0$, and the remaining
sixteen field equations, with the aid of four gauge conditions,
propagate us in time, and insure that these four are always
satisfied.}
\par
 Of course, for solutions of the
field equations we do have separately
 \beq S\^\n\_{\m;\n} =\hat S\^\n\_{\m:\n}=0,   \eeqno{gybuts}
which can be used as useful constraints of the solutions (only one
set is independent).
\par
Note the useful identities
 \beq C\_{\n}={1\over 2}\Gab g\_{\a\b}\mcd{\n}=
 -{1\over 2}\hat g\^{\a\b}{\hat g\_{\m\n}}\cd{\n}=2\k\_{,\n}/\k
 ~~~~~ \bar
 C\^{\l}=-\k^{-2}(\k^2 g\^{\l\r})\mcd{\r}~~~~C\^{\l}=-\bar
 C\^{\l}-g\^{\l\n}\mcd{\n},
 \eeqno{gequa}
 \beq \St\ten{\l}{\m\n}\cd{\l}=\hat R\_{\m\n}-{1\over 2}\hat R\_m\gmn
 -(R\_{\m\n}-{1\over 2} R\gmn)
 -(\Up\_{\m\n}-{1\over 2} \Up\gmn),
 ~~~~(C\^{\l}-\bar C\^{\l})\cd{\l}=R-\hat R\_m+\Up. \eeqno{mestu}
Identities (\ref{mestu}) follow from Eqs.(\ref{nista})(\ref{nerol}).
Similar manipulations are possible for $\Sh\ten{\l}{\m\n}$, and the
field equation (\ref{nuvec}).
\par
Contracting Eq.(\ref{misha}) with $\Gmn$ gives
 \beq  R
-[\km\M'(C\^{\l}-\bar C\^{\l})]\cd{\l} -\km\M'\Up+4\Lambda\_m =8\pi
G \T.\eeqno{mifut} Contracting Eq.(\ref{nuvec}) with $\hat
g\^{\m\n}$ gives
 \beq  \hat R-[\kp\M'({1\over 2}\bar\o\hat g\^{\l\r}
 C\_{\r}-\bar C\^{\l}
 -\hat g\^{\l\r} q\ten{\a}{\m}C\ten{\m}{\r\a})]\mcd{\l}
 +4\hat\Lambda\_m
 =8\pi G \Th,
  \eeqno{fey}
where $\bar\o=\Gmn\hat g\_{\m\n}$.  We can thus replace
Eq.(\ref{misha}) by
 \beq R\_{\m\n}
 +\km\M'(\Up\_{\m\n}
 -{1\over 2}\Up\gmn)-[\km\M'(\d\ten{\l}{\sl\m}C\_{\n\sr}
-C\ten{\l}{\m\n})]\cd{\l}+ \Lambda\_m\gmn=
 -8\pi G (\Tmn-{1\over 2}\T\gmn), \eeqno{vycool}
and similarly for Eq.(\ref{nuvec}). We can also use identities
(\ref{nista}-\ref{nerolos}) to write these equations in different
forms. Equations (\ref{mifut})(\ref{fey}) can be used to write
possibly useful integral (virial) relations by integrating them over
space-time, each with its own volume element.
\par
It was deduced in \cite{boulanger01} that under certain assumptions
about the theory, bimetric theories generically posses ghosts. One
of their assumptions was that to lowest order in departure from
double Minkowski the theory is a sum of Pauli-Fierz actions for the
different metrics, which are quadratic in the metric departures.
This, however, leads to a linear theory in this limit, which
is at odds with MOND: MOND phenomenology
dictates that at $\gmn=\hat g\_{\m\n}=\emn$ any BIMOND theory
(or any relativistic MOND theory for that matter) is not
even analytic in the squares of the departures $\gmn-\emn,~\hat g\_{\m\n}-\emn$
(where the argument of $\M'$ in the above version of the theory
vanishes, and $\M'$ diverges). It thus remains to be seen if
obstacles similar to these are at all relevant to BIMOND, and if
they are to what extent they are deleterious.
\par
For conformally related metrics $\gmn=e\^{\vartheta(x)}\hat
g\_{\m\n}$, we have
$\St\ten{\l}{\m\n}=e\^{\vartheta(x)}\Sh\ten{\l}{\m\n}$.
\subsection{Generalizations}
Some generalizations of the above simple theory include the
following.
\par
1. Instead of using $\Up$ as the argument of $\M$, we can use other
scalars, or several scalar variables. A quadratic scalar variable
can be written, most generally, as
 \beq \Xi=Q\ten{\b\c\m\n}{\a\l}C\ten{\a}{\b\c}C\ten{\l}{\m\n},
   \eeqno{cures}
where $Q\ten{\b\c\m\n}{\a\l}$ it built from $\gmn$, $\hat
g\_{\m\n}$, their inverses, $\d\ten{\a}{\b}$, and scalars such as
$\k$ and $\bar\o$. In this case the $\Lambda$ terms take a more
general form, and so do terms that are second order in the
$C\ten{\a}{\b\c}$. The only terms in $S\_{\m\n}$ and $\hat
S\_{\m\n}$ that survive in the NR limit, which we treat below, are
those involving $\St\ten{\l}{\m\n}$ and $\Sh\ten{\l}{\m\n}$. For
these we now have for example
 \beq \St\ten{\l}{\m\n}=2U\ten{\c\l}{\sl\m}g\_{\n\sr\c}
 -U\ten{\c\s}{\r}g\_{\m\c}g\_{\n\s}g\^{\l\r},~~~~~~
 U\ten{\m\n}{\l}=-2Q\ten{\b\c\m\n}{\a\l}C\ten{\a}{\b\c},
  \eeqno{tetac}
  which I shall need in what follows.
\par
For example, taking as the argument of $\M$,
$-C\ten{\c}{\m\l}C\ten{\l}{\n\c}/2\az^2$ instead of $-\Up/2\az^2$,
would leave us with only the first term in expression (\ref{gyrtaz})
for $\St\ten{\l}{\m\n}$ , and with the first three terms in
expression (\ref{tujas}) for $\Sh\ten{\l}{\m\n}$.
\par
2. One can consider more general $\a,~\b$ values.
\par
3. We can increase the symmetry with respect to the two metrics by
taking interaction terms of the form $\M(\Up\hat\Up)$,
$\M(\Up)\M(\hat\Up)$, etc..
\par
4. One can make $\M$ a function of scalars such as $\k$ and
$\bar\o$.
\par
Additional generalizations will be mentioned in section
\ref{cosmology}.

\section{\label{nonrelativistic}Nonrelativistic limit}
 Consider now the NR limit of the theory
derived from the action (\ref{gedap}). This limit applies to systems
where all quantities with the dimensions of velocities, such as $v,
~\sqrt{\f}$, etc., are much smaller than the speed of light. In the
context of GR this limit is attained by formally taking
$c\rar\infty$ everywhere in the relativistic theory. In the context
of MOND one has to be more specific, since system attributes with
the dimensions of acceleration, such as $v^2/R,~\gf$, etc., cannot
be assumed very small in the limiting process, even though they have
velocities in the numerator. We want to consider systems, such as
galaxies, in which these are finite compared with the MOND
acceleration, which is also a relevant parameter. The NR limit in
MOND is thus formally attained by taking everywhere $c\rar\infty$,
but at the same time $\ell\rar\infty$, so that $\az=c^2/\ell$
remains finite.
\par
Take a system of quasistatic (nonrelativistically moving) masses, so
that to a satisfactory approximation we can, as usual, neglect all
components of the matter EMT except $\T\_{00}=\r$. I also neglect
here the possible effects of the presence of twin
matter.\footnote{This is justified if this matter is nonexistent, or
of it is smoothly distributed so its local contribution is
negligible, or if there does not happen to exist a twin body in the
near vicinity of the $\psi$ body under study.} First, I consider the
system in a double Minkowski background. This is aesthetically the
most appealing option, which I shall assume. It relies on the
possibility that on cosmological scales the two metrics are,
somehow, maintained the same from some symmetry. There are indeed
versions of BIMOND [made more symmetric in the two metrics than our
simple action (\ref{gedap}) is] that have cosmological solutions
with $\hat g\_{\m\n}=\gmn$, either at all times, or as vacuum
solutions, which might be appropriate for today (see section
\ref{cosmology}). In this case we have $C\ten{\a}{\b\c}=0$ for the
cosmological background, and finite $C\ten{\a}{\b\c}$ values occur
only due to local inhomogeneities. We can then take locally, on
scales much smaller then cosmological ones, a double Minkowski
background. Departures from this assumption will be discussed below.
\par
Write, then, the metrics as slightly perturbed from Minkowski.
Because the source system is time-reversal symmetric in the
approximation we treat it (neglecting motions in the source), we are
looking for a solution for which the mixed space-time elements of
the two metrics vanish.\footnote{We do not have to assume this a
priori; if we do not, the equations themselves will tell us that
there is a choice of gauge in which the solution satisfies this
ansatz; see the end of this subsection. The ansatz simplifies the
presentation, and is justified a posteriori by our showing below
that such a solution exists.} We can then write most generally
 \beq \gmn=\emn-2\f\d\_{\m\n}+h\_{\m\n},~~~~
 \hat g\_{\m\n}=\emn-2\fh\d\_{\m\n}+\hat h\_{\m\n},
\eeqno{rutza} where $h\_{0\m}=h\_{\m 0}=\hat h\_{0\m}=\hat h\_{\m
0}=0$. We denote the differences
 \beq \gmn^*=\gmn-\hat g\_{\m\n}=-2\fs\d\_{\m\n}+\hs\_{\m\n},
 \eeqno{metza}
with $\fs=\f-\fh$, $\hs\_{\m\n}=h\_{\m\n}-\hat h\_{\m\n}$. We wish
to solve the field equations to first order in the potentials
$\f,~\fh,~h\_{ij}, \hat h\_{ij}$ (Roman letters are used for space
indices).
\par
Note that there is a subtlety here (as in all metric MOND theories)
due to the fact that the NR MOND potential for an isolated mass
diverges logarithmically at infinity; so, strictly speaking we
cannot formulate a first-order theory for such an isolated mass
assuming $\f\ll 1$ at all radii. However, we are, in any event,
dealing with an effective theory to be understood in the context of
the universe at large, and in this context there are no isolated
masses, our approach is meant to work only well within the distance
from the central mass to the next comparable mass, where we can
assume the first-order theory to be a good approximation.
\par
To the required order the only nonvanishing components of
$C\ten{\a}{\b\c}$ are\footnote{Because the metric derivatives,
connections, and curvature components are already first order, all
the metrics that are used to contract them can be taken as $\emn$.}
 $$ C\ten{i}{00}=C\ten{0}{0i}=C\ten{0}{i0}
 =-{1\over 2}g^*\_{00,i}=\fs\_{,i},$$
  \beq  C\ten{i}{jk}=
 {1\over 2}(g^*\_{ij,k}+g^*\_{ik,j}-g^*\_{jk,i})=
 {1\over 2}(\hs\_{ij,k}+\hs\_{ik,j}-\hs\_{jk,i})
 +\fs\_{,i}\d\_{jk}-\fs\_{,j}\d\_{ik}-\fs\_{,k}\d\_{ij}.
  \eeqno{litara}
These reflect the same relations between the separate connections
with their respective potentials.
\par
The only nonvanishing components of the Ricci tensors (shown here
for $R\_{\m\n}$) are
 \beq R\_{00}=-\D \f,~~~~~R\_{ij}
 ={1 \over 2}H\_{ij}-\D \f\d\_{ij}, \eeqno{gasum}
 with
  \beq H\_{ij}\equiv \D h\_{ij}+h\_{,i,j}-2h\_{k\sl i,j\sr,k},
  \eeqno{kuport}
where $h$ is the trace of $h\_{ij}$. The nonvanising components of
the Einstein tensor are
  \beq G\_{00}=-2\D \f+{1\over 4}H,
  ~~~~~ G\_{ij}={1\over 2}(H\_{ij}
  -{1\over 2}H\d\_{ij})\eeqno{rupdas}
($H$ is the trace of $H\_{ij}$). The same expressions exist for the
hatted and for the starred quantities. We are now ready to use these
expressions in the field equations (\ref{misha})(\ref{nuvec}). We
neglect the small cosmological-constant terms (in line with our
assuming background Minkowski metrics), and note that terms such as
$\Up\_{\m\n}$ are of second order in the potentials, so they can be
neglected. Also, $\St\ten{\l}{\m\n}$ and $\Sh\ten{\l}{\m\n}$ are
linear in components of the tensor $C\ten{\a}{\b\c}$, which are
first order in the potentials; so everywhere else in these
expressions we can take the metrics as Minkowski, so that
$f(\k)\approx 1, ~\k\_{\pm}\approx 1,~q\^{\l}\_{\m}\approx
\d\^{\l}\_{\m}$, etc.. Also, for the same reason, the covariant
derivatives can be replaced by normal derivatives. All in all we get
that the two terms involving $\St\ten{\l}{\m\n}$ and
$\Sh\ten{\l}{\m\n}$ are equal. Thus, taking the difference of the
two field equations we get
 \beq G^*\_{00}=-8\pi G \T\_{00},~~~~~~~~~~G^*\_{ij}=0.
  \eeqno{mistena}
  Substituting from Eq.(\ref{rupdas}) we get
 \beq \D \fs-{1\over 8}H^*=\fpg\r, ~~~~~~H^*\_{ij}
  -{1\over 2}H^*\d\_{ij}=0. \eeqno{jiol}
Taking the trace of the second part we get $H^*=0$, and substituting
in the first we get
 \beq \D \fs=\fpg\r. \eeqno{jiomar}
\par
We impose for $\fs$ the boundary condition at infinity $\fs\rar 0$,
which establishes it as the Newtonian potential of the problem. But
the second equation (\ref{jiol}) does not, in itself, determine
$\hs\_{ij}$, because $G^*\_{ij}$, like $G\_{ij}$, satisfy three
Bianchi identities $G^*\_{ij,j}=0$, which are the reductions of
identities (\ref{gymash}) to our case (the fourth identity is
automatically satisfied for our choice of vanishing mixed elements
of the metrics).
\par
We have only used one of the field equations (or rather their
difference). Now consider the first field equation alone in the form
(\ref{vycool}). Again, neglect the second order $\Up$ terms, etc. to
get
 \beq R\_{\m\n}+[\M'(\Sb\ten{i}{\m\n}-{1\over 2}\Sb\^{i}
 \eta\_{\m\n}
)]\_{,i}=-\fpg \r\d\_{\m\n}, \eeqno{vycas} where $\Sb\ten{i}{\m\n}$
is the NR limit of $\St\ten{\l}{\m\n}$, and $\Sb\^{i}$ its (four)
trace.  The $(0i)$ components of the equations hold identically (to
first order) since $\Sb\ten{i}{0j}=0$.\footnote{To see this note
that from Eq.(\ref{tetac}) we have $\Sb\ten{i}{0j}\propto
(2Q\ten{\b\c ji}{\a 0}-Q\ten{\b\c
 0i}{\a j})C\ten{\a}{\b\c}$.
Now, the NR limit of the $Q$ tensor is constructed only from $\emn$
and  $\Emn$. This means that its only nonvanishing components must
have three pairs of equal indices. This means, in turn, that the
only contributions to $\Sb\ten{i}{0j}$ come from $C\ten{\a}{\b\c}$
with one or three time indices; but these all vanish.}
 The $(00)$ and $(ij)$
components give, respectively,
 \beq
-\D\f+[\M'(\Sb\ten{k}{00}+{1\over 2}\Sb\^{k})]\_{,k}=-\fpg \r.
\eeqno{kuplul}
  \beq {1\over 2}H\_{ij}-\D\f\d\_{ij}+[\M'(\Sb\ten{k}{ij}
  -{1\over 2}\Sb\^{k}\d\_{ij})]\_{,k}=
 -\fpg \r\d\_{ij}.
\eeqno{kupmus} Multiply Eq.(\ref{kuplul}) by $\d\_{ij}$ and subtract
from Eq.(\ref{kupmus}) to get
  \beq {1\over 2}H\_{ij}+[\M'(\Sb\ten{k}{ij}
  -\Sb\ten{k}{mm}\d\_{ij})]\_{,k}=0, \eeqno{kuplop}
which I use instead of Eq.(\ref{kupmus}). Equation (\ref{kuplop})
does not satisfy any more identities and thus gives six independent
equations, which together with the above four unused independent
equations, and the remaining freedom to choose three gauge
conditions, should determine the remaining 13 potentials
$\f,~h\_{ij},~\hs\_{ij}$.
\par
It is beneficial to employ three of these six equations encapsuled
in Eq.(\ref{kuplop}) by taking its divergence, taking the Bianchi
identities for $H\_{ij}$ into account, to get
  \beq (\M'\Sb\ten{k}{ij})\_{,k,j}=0. \eeqno{kurda}
These three equations, together with the three independent equations
in the second of (\ref{jiol}) now involve only the six $\hs\_{ij}$
(so we managed to decouple these from $\f$ and $h\_{ij}$; $\fs$,
which also appears in these equations is already known), and can be
solved for these.\footnote{These are coupled nonlinear equations,
since $\hs\_{ij}$ appear also in the argument os $\M'$.} Once this
is done (imposing boundary conditions at infinity) $\f$ is
determined from Eq.(\ref{kuplul}) by solving a Poisson equation, and
$h\_{ij}$ are likewise determined from Eq.(\ref{kuplop}) with the
aid of gauge conditions. The remaining gauge freedom is associated
with coordinate transformations that preserve our assumed form of
the metrics; i.e., near-Minkowski and stationary (time independent
and lacking mixed elements). These are of the general form
  \beq t=t',~~~~x\^{i}=x\^{'i}+\eps\^{i}(\vx'), \eeqno{veip}
with $\eps\^{i}(\vx')$ first order in the potentials. They do not
affect $g^*\_{\m\n}$ and so leave $\hs\_{ij}$ and $\fs$
intact,\footnote{Since $g^*\_{\m\n}$ is already first order, in our
approximation the transformation affects it only to zeroth order;
i.e., not at all.} changing only $h\_{ij}$ by
$\eps\_{i,j}+\eps\_{j,i}$.
\par
Everything so far is valid for an arbitrary choice of quadratic
scalar argument. I now specialize to my preferred choice of scalar
argument $-\Up/2$. In this case we have
 \beq \Sb\ten{i}{00}=2\fs\_{,i}+{1\over 2}(\hs\_{ij,j}-\hs\_{,i}),
 ~~~~~\Sb\ten{i}{jk}={1\over 2}(\hs\_{ij,k}+\hs\_{ik,j}-\hs\_{jk,i})
+{1\over
4}[2(\hs\_{,i}-\hs\_{im,m})\d\_{jk}-\hs\_{,k}\d\_{ij}
-\hs\_{,j}\d\_{ik}].
 \eeqno{gedtup}
What is special about this case is that the space components
$\Sb\ten{k}{ij}$ depend only on $\hs\_{ij}$, not on $\fs$. This
greatly simplifies the solution of Eqs.(\ref{jiol})(\ref{kurda}),
which has to be $\hs\_{ij}=0$ (with boundary conditions
$\hs\_{ij}\rar 0$ at infinity).\footnote{All the above holds if we
use both $\Up$ and $\Up^*$ as variables, because they degenerate
into one in the NR limit. Also, note that $\bar C\^{0}=0$, and $\bar
C\^{i}=\hs\_{ij,j}-(1/2)\hs\_{,i}$; so, adding the scalar $\gmn\bar
C\^{\m}\bar C\^{n}$ should not change this conclusion.}
\par
The fact that $\hs\_{ij}=0$ causes the $C\ten{\a}{\b\c}$, as given
in Eq.(\ref{litara}), to be linear combinations of derivatives of
$\fs$, and the argument of $\M'$ becomes a function of $(\gfs)^2$.
Now Eq.(\ref{kuplul}) reads
 \beq \D\f=\fpg\r+\div\{\M'[(\gfs/\az)^2]\gfs\}. \eeqno{jirbet}
Equation (\ref{kuplop}) becomes
 \beq H\_{ij}=0, \eeqno{gushma}
and can be used with three gauge conditions, to determine
$h\_{ij}$--as would be done in GR, where this equation is always
satisfied. This implies, with the appropriate boundary conditions,
that there is a gauge in which $h\_{ij}=0$, and we work in this
gauge.
\par
The matter action for a system of slowly moving masses is, to our
present approximation,
 \beq I\_M\approx {1\over 2}\int\r(\vv^2-\f) d^3x dt ; \eeqno{juple}
So the motion of such particles is governed by the potential $\f$,
which is thus identified as the MOND potential. It is determined
from the field equation Eq.(\ref{jirbet}) (with $\fs$ being the
Newtonian potential of the system), which is the QUMOND formulation
described by Eq.(\ref{hutram}), and discussed at length in
\cite{milgrom09c}. Note that it requires solving only linear
differential equations. To have the required Newtonian limit we have
to have $\M'(z)\rar 0$ for $z\rar \infty$ (i.e., $\az\rar 0$). In
the MOND regime $z\ll 1$ we have to have $\M'(z)\approx z^{-1/4}$ to
get space-time scale invariance, which is the defining tenet of the
NR MOND limit \cite{milgrom09b} (the normalization is absorbed in
the definition of $\az$).
\par
To recapitulate, we end up with the simple result in the chosen
gauge:
 \beq \gmn=\emn-2\f\d\_{\m\n},~~~~
 \hat g\_{\m\n}=\emn-2\fh\d\_{\m\n},
\eeqno{rukun} with $\fs=\f-\fh$ being the Newtonian potential, and
the MOND potential $\f$ determined from the QUMOND
Eq.(\ref{hutram}). The relation between the first-order MOND metric
and the MOND potential is thus exactly the same as that between the
first-order GR metric and the Newtonian potential.
\par
Suppose we have not assumed a priori that the mixed elements of the
first-order metrics vanish. This does not change expressions
(\ref{litara}) for $C\ten{\a}{\b\c}$, but we now have additional
nonvanishing elements
 \beq C\ten{i}{0j}=(\hs\_{0i,j}-\hs\_{0j,i})/2,~~~~~~
C\ten{0}{ij}=-(\hs\_{0i,j}+\hs\_{0j,i})/2. \eeqno{rupaga}
 The $(0i)$
component of the difference Ricci tensor is
$R^*\_{0i}=(\D\hs\_{0i}-\hs\_{0j,i,j})/2$. So Eq.(\ref{jiol}) is now
complemented by
 \beq (\D\hs\_{0i}-\hs\_{0j,i,j})=0.  \eeqno{gumpa}
With our boundary conditions $\hs\_{0i}\rar 0$ at infinity the
solution of this equation is $\hs\_{0i}=v^*\_{,i}$ for some
$v^*(\vx)$ (the left hand side is identically divergence free, which
leaves us with only two independent equations). Note now that the
first-order limit of the theory, discussed here, enjoys a less
obvious symmetry beside the general invariance to simultaneous
coordinate transformations: it is invariant under a `small'
transformation of the form $t=t'-u(\vx')$ applied {\it separately}
to the $\gmn$ and the $\hat g\_{\m\n}$ sectors (with $u$ first order
in the potentials). In other words, there is symmetry to
transforming $g\_{0i}\rar g\_{0i}+u\_{,i}$, $\hat g\_{0i}\rar \hat
g\_{0i}+\hat u\_{,i}$ (hence $g^*\_{0i}\rar g^*\_{0i}+u^*\_{,i}$;
$u^*=u-\hat u$), with $u$ and $\hat u$ free for us to choose (the
EMTs are unchanged to lowest order). We thus have the freedom to
choose a gauge for $\hat g\_{\m\n}$ alone, in which $v^*=0$, and
hence $\hs\_{0i}=0$. This means that $C\ten{i}{0j}=C\ten{0}{ij}=0$,
and so the $(0i)$ components of Eq.(\ref{vycas}) read
 \beq (\D h\_{0i}-h\_{0j,i,j})=0,  \eeqno{gumpasa}
whose solution is $h\_{0i}=v\_{,i}$ for some $v(\vx)$. We still have
the gauge freedom to chose $v(\vx)=0$, and so we do. We thus end up
with a gauge in which the field equations themselves dictate
$h\_{0i}=\hat h\_{0i}=0$, as we assumed a priori.
\par
The double gauge symmetry we use can be seen to apply directly to
the first-order equations. It can be traced back to the fact that
the NR limit of $\Up$ is invariant to it: If we do not assume a
priori that the mixed elements $\hs\_{0i}$ vanish, then the addition
to the lowest order expression for $\Up$ is
$C\ten{i}{0k}C\ten{k}{0i}$, because $C\ten{i}{0j}$ is antisymmetric
in $i,j$, while $C\ten{0}{ij}$ is symmetric. But, under the double
gauge transformation $\hs\_{0i}$ changes by $u^*\_{,i}$, so
$C\ten{i}{0k}$ is invariant, and so is $\Up$.
\par
It remains to be checked if this symmetry is a remnant of some
symmetry enjoyed by the relativistic theory itself.
\par
Anticipating the discussion of the next subsection, note that not
all scalar arguments are invariant to this double gauge in their NR
limit. For example, the change induced in the scalar $\gmn\bar
C\^{\m}\bar C\^{\n}$ is $-\hs\_{0i,i}\D u^*$. For the more general
case, Eq.(\ref{gumpa}) is still valid, and again gives
$\hs\_{0i}=v^*\_{,i}$, while instead of Eq.(\ref{gumpasa}) we have
more generally
 \beq (\D h\_{0i}-h\_{0j,j,i})/2+[\M'\Sb\ten{k}{0i}]\_{,k}=0.
 \eeqno{nukata}
The first term is identically divergence free so we can write one of
these three equations as
 \beq [\M'\Sb\ten{k}{0i}]\_{,k,i}=0.  \eeqno{milorat}
Now, $\Sb\ten{k}{0i}$ is linear in $v^*$ (which may also appear in
the argument of $\M'$), so this equation generically dictates
$v^*=0$. For the scalar $\Up$ this does not work because at this
stage we already have $\Sb\ten{k}{0i}=0$, but in return we have the
double gauge freedom to help us remove $v^*$. For the scalar
$\gmn\bar C\^{\m}\bar C\^{\n}$ we do not have the double gauge
freedom but we can write $\Sb\ten{k}{0i}\propto \d\^k\_i\D v^*$, so
Eq.(\ref{milorat}) gives $\D(\M'\D v^*)=0$, which implies $v^*=0$.
In any event we can always have $\hs\_{0i}=0$, and continue from
there as before to show that there is always a gauge where the mixed
elements of the metrics vanish.

\subsection{Other choices of the scalar argument of $\M$}
 Here I consider the NR limit
of theories with other choices of the quadratic scalar argument of
$\M$. The main purpose is to see whether these give theories that
are different in their NR limits, and so can be distinguished using
observations of NR systems such as rotation curves and lensing in
galaxies.
\par
Take then the general quadratic argument as given by
Eq.(\ref{cures}). All of our procedures in the present section, up
to Eq.(\ref{veip}) remain valid. Up to that point we had not made
use of the particular expressions for $\St\ten{\l}{\m\n}$ and
$\Sh\ten{\l}{\m\n}$, only of the fact that they become equal to
first order in the potentials, and this is still the
case.\footnote{This follows from the asymmetry of $C\ten{\a}{b\c}$
to interchange of the matrices, and from the fact that to first
order we can put everywhere else $\hat
g\_{\m\n}\approx\gmn\approx\emn$.}
\par
Departure from the above occurs, however, for more general scalars
in the employment of Eq.(\ref{kurda}). Now, the space components
$\Sb\ten{i}{jk}$ do, in general, depend on the gradient of $\fs$,
and it is easy to see that $\hs\_{ij}=0$ is no longer a solution, in
general: Substituting $\hs\_{ij}=0$ in Eq.(\ref{kurda}) would result
in three constraints on the Newtonian potential $\fs$, which it does
not satisfy ($\fs$ is really arbitrary if we allow arbitrary density
distributions, including negative ones). So, in general, after $\fs$
is calculated as the Newtonian potential of the system, we have to
solve the nine coupled equations (\ref{jiol}) (second part) and
(\ref{kurda}) (only six of which are independent) for the
$\hs\_{ij}$. After this is done we determine $\f$ from
Eq.(\ref{kuplul}), and $h\_{ij}$ from Eq.(\ref{kuplop}) with the aid
of gauge conditions. Note that in the Newtonian limit, $\az\rar 0$,
$\f$ becomes the Newtonian potential, and $h\_{ij}\rar 0$ as fast as
$\M'$ does.
\par
We can derive some scaling properties of the $\hs\_{ij}$, even in
the general case: It is easy to see that the second  Eq.(\ref{jiol})
and Eq.(\ref{kurda}) are invariant under $m_i\rar \l m_i,~\vr\rar
\l^{1/2}\vr$, where $m_i$ stand for the masses in the system. This
is because the only quantities appearing in these equations, beside
the variables $\hs\_{ij}$, are $\fs\_{,i}$ and $\az$ both of the
same dimensions as $m_iG/r^2$. This means that the $\hs\_{ij,k}$ are
invariant under this scaling.
\par
Consider now in more detail a spherically symmetric problem. We are
looking for solutions in which the various potential tensors
$h\_{ij},~\hat h\_{ij},~\hs\_{ij}$ are of the form exemplified by
 \beq \hs\_{ij}=\hs\_1(r)\d\_{ij}+\hs\_2(r)n\_{i}n\_{j},
  \eeqno{cuplut}
where $\vn=\vr/r$. It can be shown that $\hs\_{ij}$ satisfying the
second Eq.(\ref{jiol}), namely, annuling $H^*\_{ij}$, is tantamount
to $\hs\_2=r{\hs\_1}'$, which means, in turn, that
$\hs\_{ij}=q\_{,i,j}$ for some $q(r)$. It thus remains to determine
$q$ from Eq.(\ref{kurda}). Because of the spherical symmetry, the
different $i$ components of Eq.(\ref{kurda}) give equivalent
equations;\footnote{This equation has to read in the spherical case
$P[q(r)]\vr=0$ where $P$ is a differential operator acting on
$q(r)$, and we get one equation $P[q(r)]=0$.} so we are left with
only one equation from which to determine $q(r)$.
\par
Once $q(r)$ is known, we use Eq.(\ref{kuplop}) to solve for
$h\_{ij}$. Write $h\_{ij}$ in the form (\ref{cuplut}). We can use
the remaining gauge freedom to eliminate one of the two functions.
In the spherically symmetric case the remaining freedom is to
transform $\vr=\vr'[1+\eps(r')]$ for some $\eps(r')$, treated to
first order. This transformation takes $h\_{ij}\rar
h\_{ij}+2\eps\d\_{ij}+2r\eps' n\_i n\_j$; so, we can use such a
transformation to eliminate either of the functions in the
expression for $h\_{ij}$. For example, let us choose the gauge in
which
 \beq h\_{ij}= \varphi(r)\d\_{ij}.  \eeqno{jolta}
The general NR MOND metric is thus diagonal for this choice of gauge
with
 \beq g\_{00}=-1-2\f,~~~~~g\_{ij}=\d\_{ij}[1-2(\f+\varphi)].
  \eeqno{tamga}
 For the form (\ref{jolta}) of $h\_{ij}$ we have
  \beq H\_{ij}-{1\over 2}H\d\_{ij}=-(\varphi''
  +r^{-1}\varphi')\d\_{ij}+(\varphi''-r^{-1}\varphi')n\_i n\_j\equiv
a\d\_{ij}+b n\_i n\_j.  \eeqno{jumba}
  Note that $a'=-r^{-2}(r^2b)'$, from the fact that the
expression is divergence free.
 We now use
  \beq H\_{ij}-{1\over 2}H\d\_{ij}=-2(\M'\Sb\ten{k}{ij})\_{,k},
  \eeqno{kurkas}
obtained from Eq.(\ref{kuplop}), to solve for $\varphi$. Since the
right hand side of this equation is already known to be divergence
free, from Eq.(\ref{kurda}), we get only one independent equation of
the form
  \beq r(r^{-1}\varphi')'=p(r),  \eeqno{soter}
where $p(r)$ is a known function. This we finally solve for
$\varphi$, which we permit to behave asymptotically as $ln(r)$.
\par
Here we note already an interesting difference from the theories
that have $\Up$ as scalar argument, which  have Eqs.(\ref{hutred})
and (\ref{hutram}) as their NR limit (even with general
$\a,~\b$--see the next subsection). In such theories, in the
spherical case the MOND acceleration is an algebraic function of the
Newtonian one, with the relation being unique for the theory. In the
general case this is not so: To get the MOND acceleration in the
spherical case we apply the Gauss theorem to Eq.(\ref{kuplul}). The
expression we then get is some functional of $q(r)$ that cannot be
written as a function of the Newtonian acceleration $-d\fs/dr$. This
can lead to different predictions even for massive-particle motions.
\par
The spherical problem can be solved analytically for the case where
$\fs=Ar^\t$--for example when we are outside the mass where $\t=-1$,
and we are in a region where $M'(z)\propto z^{-\sigma }$, for
example, in the deep-MOND regime where we will have to have $\sigma
=1/4$. Then the solution can be shown to be of the form $q=\l
r^2\fs$, with $\l$ determined from Eq.(\ref{kurda}) depending on
$\t$ and $\sigma $. With this ansatz $\M'\propto
(|A|/\az)\^{-2\sigma }(a+b\l+c\l^2)^{-\sigma }r^{-2\sigma (\t-1)}$,
and Eq.(\ref{kurda}) then gives $(A/\az)\^{-2\sigma
}(a+b\l+c\l^2)^{-\sigma }A(\bar a +\bar b\l)r^\z\vn=0$, with
$\z=(1-2\sigma )(\t-1)-2$ ($=-3$ for the above examples), and $\bar
a,~\bar b$ depending on $\t,~\sigma $ and the choice of scalar
argument of $\M$ for the specific theory ($\bar a$ comes from the
terms in $\Sb\ten{k}{ij}$ linear in the gradient of $\fs$, and $\bar
b$ from those linear in the gradient of $\hs\_{ij}$). So $\l=-\bar
a/\bar b$ gives us the solution (for the choice of $\Up$ as argument
$\bar a=0$). Equation (\ref{soter}) then gives $\varphi=\xi
(A/\az)^{-2\sigma }Ar\^{\z+3}$ for $\z\not = -3$, and $\varphi=\xi
(A/\az)^{-2\sigma }Aln(r)$ for $\z= -3$, with the dimensionless
$\xi$ determined.
\par
Take, for instance the interesting case where we are asymptotically
outside matter and in the deep-MOND regime, where $A=-MG$ and
$\z=-3$. We then get $\varphi= \xi (MG\az)^{1/2}ln(r)$,
asymptotically. In this case we also have $\f= (MG\az)^{1/2}ln(r)$,
by definition; so,  $\varphi=\xi\f$. For example, for the choice of
argument $-\Gmn C\ten{\c}{\m\l}C\ten{\l}{\n\c}/2\az^2$, I find,
following the above procedure for the deep-MOND ($\s=1/4$),
asymptotic ($\t=-1$) case: $\xi=4/3$.
\par
In summary, the asymptotic form of the MOND metric is diagonal, with
 \beq g\_{00}=-1-2\f,~~~~~g\_{ij}=\d\_{ij}[1-2\f(1+\xi)].
  \eeqno{sumga}
\par
Remember that while $\varphi=\xi\f$ in the MOND regime, in the
Newtonian regime $\f$ becomes the Newtonian potential while
$\varphi$ vanishes as fast as dictated by the vanishing of $\M'$ at
high values of its argument.
\subsection{General $\a~\b$ values}
The NR limit of the field equation in a theory governed by the
action (\ref{gedat}) for general $\a$ and $\b$ values is
 \beq \b(R\_{\m\n}-{1\over 2}\emn R)
 +(\M'\Sb\ten{i}{\m\n})\_{,i}=-8\pi G \r\d\_{\m 0}\d\_{\n 0},
  \eeqno{cutek}
 \beq \a(\hat R\_{\m\n}-{1\over 2}\emn \hat R)
 -(\M'\Sb\ten{i}{\m\n})\_{,i}=0,
  \eeqno{cutan}
with all quantities taken to first order in the potentials $\f$,
$\fh$, $h\_{ij}$, $\hat h\_{ij}$. $\Sb\ten{i}{\m\n}$ is linear in
the first-order expression for $C\ten{\a}{\b\c}$, as given in
Eq.(\ref{litara}), and  $\M'$ is a function of a quadratic scalar
built from these. Multiply the second equation by $-\b/\a$ and add
to the first to get
 \beq \b(R^*\_{\m\n}-{1\over 2}\emn R^*)
 +{\a+\b\over \a}(\M'\Sb\ten{i}{\m\n})\_{,i}
 =-8\pi G \r\d\_{\m 0}\d\_{\n 0}.
  \eeqno{cutaya}
The $(0j)$ components of this equation hold identically, as before.
Equations (\ref{gasum})-(\ref{rupdas}) are then used to write the
$(00)$ and $(ij)$ components of this equation as
 \beq \b(\D \fs-{1\over
8}H^*)-{\a+\b\over 2\a}(\M'\Sb\ten{i}{00})\_{,i}=\fpg\r,
~~~~~~H^*\_{ij}
  -{1\over 2}H^*\d\_{ij}
  +{2(\a+\b)\over \a\b}(\M'\Sb\ten{k}{ij})\_{,k}=0. \eeqno{jopas}
Taking the trace of the second and substituting for $H^*$ in the
first we get
  \beq \b\D \fs-{\a+\b\over 2\a}[\M'(\Sb\ten{i}{00}
  +\Sb\ten{i}{mm})]\_{,i}=\fpg\r.
\eeqno{gruda} Equation (\ref{jopas}) comprises seven independent
equations that can be solved for $\fs$ and the $\hs\_{ij}$. Once
these are known we can get $\f$ and $h\_{ij}$ from Eq.(\ref{cutek}),
or equivalently from
 \beq \b R\_{\m\n}
 +[\M'(\Sb\ten{i}{\m\n}
 -{1\over 2}\Sb\^{i}\emn)]\_{,i}=-\fpg \r\d\_{\m\n}.  \eeqno{cutata}
 Its $(00)$ component gives
 \beq \D\f=\fpg\b^{-1}\r+\b^{-1}[{1\over 2}\M'(\Sb\ten{i}{00}
 +\Sb\ten{i}{kk})]\_{,i} \eeqno{hutves}
from which $\f$ can be gotten (since the right hand side is now
known). The $(ij)$ component gives [after making use of
Eq.(\ref{hutves})]
 \beq H\_{ij}-{1\over 2}H\d\_{ij}=-2\b^{-1}(\M'\Sb\ten{k}{ij})\_{,k}.
\eeqno{nuref} From these $h\_{ij}$ can be gotten. Note that the
divergence of Eq.(\ref{nuref}) is identically the same as that of
the second Eq.(\ref{jopas}); so we only have here three independent
equations for the six $h\_{ij}$. However, again we have the gauge
freedom to eliminate this indeterminacy.
\par
Specializing to our preferred choice of scalar $\Xi=-\Up/2$ the NR
limit of the theory again greatly simplifies since from
Eq.(\ref{gedtup}) $\Sb\ten{k}{ij}$ does not contain the derivatives
of $\fs$. This implies that the second term in the second
Eq.(\ref{jopas}) is linear in the derivatives of $\hs\_{ij}$, and so
the solution of this equation is easy to get: $\hs\_{ij}=0$, (again,
with the asymptotic boundary conditions $\hs\_{ij}\rar 0$). This
means that the argument of $\M'$ is now a function of
$(\gfs/\az)^2$, and that, in fact, $\Sb\ten{i}{jk}=0$. As a result
the first of Eq.(\ref{jopas}) becomes identical with the first of
Eq.(\ref{hutred}), while Eq.(\ref{hutves}) becomes identical with
the second of Eq.(\ref{hutred}). In addition, from Eq.(\ref{nuref})
we get $h\_{ij}=0$ as before.
\par
We thus end up with a NR limit in which $\gmn=\emn-2\f\d\_{\m\n}$
and $\hat g\_{\m\n}=\emn-2\fh\d\_{\m\n}$ as in Eq.(\ref{rukun}),
with the MOND potential $\f$ determined now from the NR  MOND theory
described by Eq.(\ref{hutred}). This theory has been discussed at
length in \cite{milgrom09c}. The relation between the first-order
MOND metric and the MOND potential is thus, again, the same as that
between the first-order GR metric and the Newtonian potential.
\par
Note in this context as well that using a scalar argument that is a
combination of $\Up$ and $\bar\Up=\gmn\bar C\^{\m}\bar C\^{\n}$,
leads to the same first-order metric and NR limit.
\subsection{Other backgrounds}
So far I assumed that the two metrics have the same cosmological
background and so, for systems small on the cosmological scale both
can be taken as nearly Minkowski.
\par
Here I consider some possible departures from this assumption. One
possibility, for example, is that for today's cosmology the two
metrics are conformally related $\hat g\_{\m\n}=\oo \gmn$ with
constant $\oo$; this still gives $C\ten{\a}{\b\c}=0$ for the
cosmological background. In this case we can take locally the
background metrics to be $\gmn\^{B}=\emn,~\hat
g\_{\m\n}\^{B}=\oo\emn$ and expand around these:
  \beq \gmn=\emn-2\f\d\_{\m\n}+h\_{\m\n},~~~~
 \hat g\_{\m\n}=\oo(\emn-2\fh\d\_{\m\n}+\hat h\_{\m\n}),
\eeqno{ramsha} instead of Eq.(\ref{rutza}). The expressions of
$\hat\C\ten{\a}{\b\c}$, $C\ten{\a}{\b\c}$, and $\hat R\_{\m\n}$ in
terms of the potentials remain the same as before. Also, `small'
coordinate transformations of the type shown in Eq.(\ref{veip})
still do not affect the potential differences only the $h\_{ij}$.
Work with the general theory for arbitrary $\a,~\b$. The NR limit of
the field equations is now
  \beq \b(R\_{\m\n}-{1\over 2}\emn R)
 +\oo f(\oo^{-1})(\M'\Sb\ten{i}{\m\n})\_{,i}=-8\pi G \Tmn,
  \eeqno{pusek}
 \beq \a(\hat R\_{\m\n}-{1\over 2}\emn \hat R)
 -f(\oo^{-1})(\M'\Sb\ten{i}{\m\n})\_{,i}=0,
  \eeqno{cpusan}
with the $\oo$ powers coming from factors such as $\k\_{\pm}$, etc..
Defining $\tilde\a=\oo\a$, and $\tilde\M'=\oo f(\oo^{-1})\M'$, we
get back the $\oo=1$ case, but with $\a$ replaced by $\tilde\a$, and
$\M'$ by $\tilde\M'$. As before, with our favorite choice of scalar
argument $\Xi=-\Up/2\az$ we have $h\_{ij}=0$ in the appropriate
gauge; so we get the relation between the MOND metric and the MOND
potential as before.
\par
For a more general background we can write the background
metrics,locally for a small system,
 \beq \gmn=\emn,~~~~~\hat g\_{\m\n}=\oo(\emn-u\d\_{\m 0}\d\_{\n 0}).
 \eeqno{goshet}
This leads to more complex NR limits, which I do not discuss here.
\par
We see then that the NR limit of the theory as applied to system
that are small on cosmic scale depends on the background metrics. If
the relation between the background metrics vary with cosmological
time (I assume that it does not--see section \ref{cosmology}) the
application of BIMOND to local inhomogeneities also varies with
cosmic time. But I will not discuss this possibility further here.

\section{\label{gr}General Relativity limit}
 I showed in \cite{milgrom09c} that the NR theories with
$\b=1$, but arbitrary $\a$, have a Newtonian limit if $\M'(z)\rar 0$
for $z\rar\infty$. This carries over to the relativistic theories.
The same property of $\M$ causes the relativistic theory to go to GR
in the same limit $\az\rar 0$, because $\M'\rar 0$ implies $\tilde
T\_{\m\n},\hat T\_{\m\n}\rar 0$, and since in this case $\M(z)/z$
must also vanish in the limit we also have
$\Lambda\_m,~\hat\Lambda\_m\rar 0$. This gives GR, with $\gmn$
satisfying the Einstein equation with the standard matter EMT as
source. In this limit $\hat g\_{\m\n}$ satisfies its own Einstein
equation. It decouples from $\gmn$ anyway, so it affects matter
neither directly nor indirectly.
\par
We do not have tight phenomenological constraints on how fast MOND
approaches Newtonian dynamics for high accelerations. There are
indications from solar system constraints \cite{milgrom83,
milgrom09a,sj06} that $\M'(z)$ decreases at least as fast as
$z^{-1}$, but it might turn out to do so much more precipitously. If
so, any departure from GR might be practically wiped out in a
very-high-acceleration system, such as a laboratory on earth, the
inner solar system, or a close binary pulsar.\footnote{The presence
of the galactic field still induces the departures discussed in
\cite{milgrom09a}.}$^,$\footnote{A theory like TeVeS also has a
surrogate for $\az$ that appears in its NR limit, which is
constructed out of the constants characterizing the theory. However,
TeVeS does not become GR in the limit $\az\rar 0$ and this leaves
possibly detected effects even in an isolated solar system, the
binary pulsar, etc., even with the very high accelerations
characterizing them
\cite{bek04,sanders05,bruneton07,sagi09,skordis09}}
\par
In the limit $\az\rar 0$, the vanishing of $\tilde T\_{\m\n}$ and
$\hat T\_{\m\n}$ might occur much faster than that of $\Lambda\_m$
and $\hat\Lambda\_m$. For example, if $\M(\infty)$ is finite, the
latter vanish as $\az^2$, wile the former might vanish much faster.
We can thus, as an intermediate approximation, keep the $\Lambda\_m$
and $\hat\Lambda\_m$ terms in the theory, and write the limiting
field equations (\ref{misha})(\ref{nuvec}) as
  \beq  R\_{\m\n}-{1\over 2}R\gmn -\Lambda\_m(\infty)\gmn=
 -8\pi G \Tmn,  \eeqno{misva}
 \beq  \hat R\_{\m\n}-{1\over 2} \hat R\hat g\_{\m\n}
-\hat\Lambda\_m(\infty)\hat g\_{\m\n} = -8\pi G \hTmn. \eeqno{nulat}
\par
For general $\b$ values it remains to be checked whether the
requirement on $\M'$ from the NR limit, deduced in \cite{milgrom09c}
suffices to guarantee a GR limit for $\az\rar 0$.

\section{\label{lensing}Gravitational lensing}
 One of the two main phenomenological duties we
expect from a relativistic MOND theory is to predict gravitational
lensing correctly. In particular, we know that lensing analysis of
galactic systems  indicate mass discrepancies that are not very
different to those derived from massive-particle motions (e.g.,
rotation curves). In other words, lensing in the MOND regime is
found observationally to be greatly enhanced over the GR prediction
without DM. Reproducing this fact has been a pressing desideratum in
constructing relativistic MOND theories, achieved finally in TeVeS
through the efforts of Sanders \cite{sanders97} and Bekenstein
\cite{bek04}.
\par
In the present BIMOND class such enhanced, MOND-like lensing is
predicted naturally by all the theories in the class. For a choice
of the scalar argument that is a combinations of $\Up$, $\Up^*$, and
$\bar\Up$ (and with any $\a,~\b$), relation (\ref{rukun}) between
the first-order MOND metric and the MOND potential holds. So, to
this order the MOND connection $\C\ten{\l}{\m\n}$ is expressed in
terms of the MOND potential $\f$ in the same way as the GR
connection is expressed in terms of the Newtonian potential. This
means, in turn, that the MOND potential describes the dynamics of
both massive and massless particles in the same way as the Newtonian
potential does in GR. In other words, such theories predict that
analyzing lensing and massive-particle dynamics by a NR system
assuming GR, should give the same effective potential (or the same
distribution of ``phantom matter''). This is consistent with
observations.
\par
We also saw that there are other choices of the scalar argument of
$\M$ for which the NR MOND metric is characterized by additional
potentials $h\_{ij}$. However, these vanish quickly for
accelerations much above $\az$, while  in the MOND regime they are
of the same order as the MOND potential $\f$. We then expect these
theories to yield somewhat different lensing to that expected with
the GR relation between metric and potential, albeit still with the
MOND characteristics.
\par
For example, we saw that far from a central mass $M$, and in the
deep-MOND regime, the form of the MOND metric is given by
Eq.(\ref{sumga}), and this leads to lensing that is multiplied by a
factor $1+\xi/2$ over that expected from the MOND potential with the
GR prescription.
\par
Eventually, by comparing lensing and massive-particle dynamics in
the low-acceleration fields of galaxies or other galactic systems,
we may be able to differentiate observationally between theories
with different scalar arguments.

\section{\label{cosmology}Cosmology}
 I cannot at present offer a specific BIMOND
cosmology. There are two obstacles to doing this. In the first place
we cannot even pinpoint the exact BIMOND theory out of the various
versions possible. NR phenomenology can assist somewhat in
pinpointing the NR limit. However, even for a given NR MOND theory
there are different relativistic versions having this limit, which
differ greatly in the relativistic regime, and specifically in their
application to cosmology.
\par
Second, as we well know from a century of experience with GR, having
the underlying theory is one thing; pinpointing the cosmology is
another: In dealing with standard GR cosmology, the cosmological
evolution and the present state of the universe does not emerge
uniquely from first principles. There are major observational
constraints, assumptions about symmetries, initial conditions, and
matter content, that are put in by hand into cosmological theory. We
do not know, for instance, the initial conditions for our universe
from first principles, so an initial singularity (as opposed say to
a steady state universe with continuous matter creation, or to a
static universe as Einstein would have it initially) is imposed by
hand. Early inflation (the mechanism for which is still moot) is put
in by hand to account for various observational facts. Cosmic
acceleration, whose cause remains unknown, is imposed by hand by
invoking `dark energy', modified gravity, or other mechanisms. The
material content of the universe (e.g. the very existence of baryon
asymmetry--mechanism still unknown) is an input in cosmology. A
priori, we could have had a cosmos with a space that is
inhomogeneous or anisotropic on cosmological scales; but, the
cosmological principle is imposed based on what our eyes tell us
about our universe. All this is even more acute in light of recent
developments in quantum cosmology.
\par
In the case of BIMOND we are on even shakier ground when coming to
construct a cosmology. Here we are dealing with two space-times,
only one of which we can sense directly. We have no direct knowledge
of many of the global properties of the other space-time. Did it
have a Big Bang? Did it undergo inflation? Is it spatially flat? (in
Rosen's theory the auxiliary metric is constrained to be flat)? Is
it spatially homogeneous and isotropic? What is the nature of the
twin matter? Is it there at all? Is it always homogeneously
distributed, or does it clump? Is it characterized by the same
baryon asymmetry, etc.?
\par
MOND phenomenology in systems that are small on cosmological scales
is particularly simple and clear cut in a double Minkowski
background. Such a background applies if in cosmology $\hat
g\_{\m\n}=\gmn$. BIMOND cosmology that have disparate metrics, so
that the cosmological values of $C\ten{\a}{\b\c}$ are appreciable,
do not seem to make phenomenological sense in small systems.

I thus assume, as an additional cosmological assumption to the many
above, that on cosmological scales we have $\hat g\_{\m\n}=\gmn$.
This could emerge, as an external constraint on BIMOND, from the
world picture that underlies it. Or, more appealingly, it could, at
least, correspond to a solution of BIMOND itself in some version.
\par
I thus consider here briefly only cosmologies with $\hat
g\_{\m\n}=\gmn$, or with the somewhat relaxed assumption $\hat
g\_{\m\n}=\oo\gmn$, with a constant $\oo$. In either case we have
$C\ten{\a}{\b\c}=0$ in cosmology; so, finite values of
$C\ten{\a}{\b\c}$ are produced only due to local inhomogeneities.
This greatly simplifies the equations of motion, since the only
contributions of the interaction term that survive are the
$\Lambda\_m$ terms. With this ansatz, the equations of motion for
the more general action (\ref{gedat}) are then
 $$ \b G\_{\m\n}+q\az^2\M(0)\gmn
 =-8\pi G\Tmn(\gmn,\psi_i),$$
 \beq \a  G\_{\m\n}+\hat q\az^2\M(0)\gmn
 =-8\pi G\hTmn(\oo\gmn,\chi_i), \eeqno{juopl}
where
 \beq q=(\oo/2)[\k f(\k)]'\_{\k=\oo^{-1}}, ~~~~
\hat q= -(\oo^{-2}/2)[\k^{-1} f(\k)]'\_{\k
 =\oo^{-1}}, \eeqno{lures}
and I used the fact that with the above ansatz $\hat
G\_{\m\n}=G\_{\m\n}$. For the two equalities in equation
(\ref{juopl}) to hold simultaneously, we need to start with a BIMOND
theory with some symmetry with respect to the two metrics. As an
example consider a special, more symmetric, case of the
gravitational Lagrangian in Eq.(\ref{gedat}) written as
 \beq  \b \gh (R-2\az^2\bar \M) +\a\hgh (\hat R-2\az^2\bar \M).
  \eeqno{gutsga}
 This choice corresponds to $f(\k)=(\b\k+\a\k^{-1})/(\a+\b)$, and
$\M=(\a+\b)\bar \M$. It gives $q=\b/(\a+\b)$, $\hat
q=\oo\a/(\a+\b)$. So the field equations are now
  $$ \ G\_{\m\n}+\az^2\bar\M(0)\gmn
 =-8\pi G\b^{-1}\Tmn(\gmn,\psi_i),$$
 \beq G\_{\m\n}+\l\az^2\bar\M(0)\gmn
 =-8\pi G\a^{-1}\hTmn(\oo\gmn,\chi_i). \eeqno{jural}
Clearly, $\hat g\_{\m\n}=\gmn$ ($\oo=1$) always corresponds to a
vacuum solution of this theory, with both space-times being a de
Sitter or Anti de Sitter, with a cosmological constant $\Lambda
=-\az^2\bar\M(0)$. Furthermore, if we also have from symmetry, for
two identical configurations in the two sectors,
$\hTmn=(\a/\b)\Tmn$,\footnote{This ensures that without the
interaction, physics is the same in the two sectors.} then the two
equations are the same even with matter. The cosmology we then get
is, quite interestingly, standard GR cosmology (taking $\b=1$) with
$\Lambda$ as cosmological constant. This would be reassuring since
it would automatically insure that we are not bereft of the
successes of standard cosmology regarding inflation,
nucleosynthesis, etc.\footnote{This line of thinking seems to
indicate that the theories with $\a=\b=1$ are preferable.}
\par
This picture would also force us to consider more seriously the
nature of the twin matter, and its possible visible effects in our
space-time. If it is homogeneously distributed it will be difficult
to detect any direct effects of it. If it clumps it could have
various effects; for example, it may produce some effects that are
otherwise attributed to cosmological dark matter.
\par
It is also possible that BIMOND can replace cosmological DM by the
distribution and fluctuations in $\tilde T\_{\m\n}$, which is
constructed from the two metrics alone, not directly from matter.
This would be similar in vein to what has been discussed in
connection with such a possible role of auxiliary fields in other
theories \cite{sanders05,z08,skordis09,banados09}.
\par
The above is only one example of a BIMOND theory that has a
cosmology with $\hat g\_{\m\n}=\gmn$. This particular version should
not be assumed for the special case $\b+\a=0$, since then $\bar\M$
drops from the theory in the NR limit. However there are other
versions of BIMOND that can accommodate our cosmological ansatz for
this case. For example, we can take as the gravitational Lagrangian
density (say with $\b=1$)
 $$ \gh (R-2p\az^2)
  -\hgh (\hat R-2\hat p\az^2)-2(g\hat g)^{1/4}f(\k)\az^2\M=$$
 \beq  =\gh R
  -\hgh \hat R-2\az^2(g\hat g)^{1/4}[p\k-\hat p \k^{-1}+f(\k)\M],
  \eeqno{gukat}
This theory also has the NR limit we discussed above, and has a
cosmological solution with $\hat g\_{\m\n}=\gmn$ if a certain
relation between $p,~\hat p$, and $\M(0)$ holds. For example, if
$\M(0)=0$ we have the symmetric theory with $p=\hat p$, in which
case $-p\az^2$ is the cosmological constant. More generally, we can
take the gravitational Lagrangian density (keeping the symmetry)
  $$ \gh (R-2\bar\M\az^2)
  -\hgh (\hat R-2\bar\M\az^2)-2(g\hat g)^{1/4}f(\k)\az^2\M=$$
  \beq  =\gh R
  -\hgh \hat R-2\az^2(g\hat g)^{1/4}[(\k- \k^{-1})\bar\M+f(\k)\M]
  \eeqno{gupal}
(with $\bar\M$ also a function of the scalar argument $\Up/\az^2$).
Then, $\bar\M$ does not appear in the NR limit on a double Minkowski
background, which is governed by $\M$, and a cosmology with our
ansatz is a solution, with $-\az^2\bar\M(0)$ being the cosmological
constant [for $\M(0)=0$].
\par
We saw that in the GR limit $\az^2\M\_{\infty}$
 plays the role of a cosmological constant [$\M\_{\infty}=\M(\infty)$],
while in the present context it is $\az^2\M(0)$.  More generally,
the $\Lambda\_m\propto \az^2\M$ term, which is variable, and
possibly other terms in $\tilde T\_{\m\n}$, may give rise to ``dark
energy'' effects. $\M(z)$ need not change much for the full range of
$z$, since for small values it also has to go to some constant, as
$\M(z)\approx \M_0+(4/3)z^{3/4}$ for $z\ll 1$. In fact, changes in
$\M(z)$ over the full $z$ range are, generically, of order unity,
since $\M\_{\infty}-\M(z)=\int\_z\^{\infty}\M'(z)dz$ is of order
unity if $\M'$ decreases beyond $z=1$ faster than $z^{-1}$ (unlike
$\M$, which is known only up to an additive constant, $\M'$ is
determined by MOND phenomenology). $\M\_{\infty}$ is a dimensionless
constant characterizing the theory. If $|\M\_{\infty}|\sim 1$, then
$|\M|$ is always of order unity. We then automatically get the well
known, but otherwise mysterious, proximity between $\az^2$, as
determined from the dynamics of small systems, and $\Lambda$, the
density of `dark energy', as deduced from cosmology.\footnote{The
possible proximity of $\Lambda$ and $\az^2$, thus hinges on the
dimensionless $\M$ being of order unity. Because of the way the
normalization of $\M$ is defined, this means that the scale over
which $\M$ varies as a function of $\Up$, and the scale that
determines the magnitude of the $\M$ term in the Lagrangian, which
have the same dimensions of length$^{-2}$, are also of the same
magnitude. This need not be the case, just as not all mass
parameters that appear in the standard model of particle physics are
of similar values. So, the apparent proximity $\Lambda\sim\az^2$
that we get here is only a plausibility not a corollary.}
\par
Note that irrespective of the relevance to MOND, bimetric theories
of the type presented here provide a frame for discussing `dark
energy' as modified gravity, which could be an alternative to
schemes such as $f(R)$ theories.
\subsection{Deep-MOND relativistic systems?}
In principle, BIMOND theories enable one to study the structure of
deep-MOND, relativistic systems such as deep-MOND black holes. As
has been stressed many times in the past, such a deep-MOND system
would have to have its typical curvature radius much larger than the
MOND length $\ell=c^2/\az$. This length is, however, of the order of
the Hubble radius today, and certainly in the past. In practice
then, the universe seems to be the only such low-acceleration
(rather, intermediate-acceleration) relativistic system, at present.
\section{\label{discussion}Discussion}
I have  described  a class of bimetric MOND (BIMOND) theories.
Matter lives in  the space-time described by one of the metrics,
which, in turn, couples to another through the interaction $\M$
term. If we heuristically view gravity as reflecting an effective
`elasticity' of space-time we can view the double-metric nature of
our theory as representing two coexisting elastic bodies, each with
its own elasticity as encapsuled in the respective $R$, $\hat R$
terms in the action. Thus MOND departure from GR is introduced not
through a modification of the `elasticity' properties of space-time,
but rather through the interaction of the space-time that is the
arena for matter with the auxiliary one. The strength of the
interaction between these two space-time `membranes' depends on the
gradient difference. The response of our home space-time to matter
is affected by its interaction with the other space-time, which
modifies its effective elasticity. However, once its shape is
determined, this home space-time affects matter in the standard way.
With our assumption that on cosmological scales $\hat
g\_{\m\n}=\gmn$ the two membranes are, in a sense, stuck together on
these scales, and `separate' only locally due to inhomogeneities.
Such heuristics may help pinpoint the fundamental concept
underpinning the MOND paradigm. For example, it may give meaning to
the length $\ell=c^2/\az$ that appears in the NR limit as $\az$.
\par
The BIMOND theories have the (yet unproven) potential to account for
all the components of the dark sector (galactic DM, cosmological DM,
and dark energy) from one term in the action, all controlled by
$\az$.
\par
My main objective has been to point out that there exists such a
class of relativistic theories that have MOND-like theories as their
NR limit, and, which produce enhanced, MOND-like gravitational
lensing. We are, however, still far from pinpointing the exact
version of the theory that is the most suitable. This is
particularly true in the context of cosmology, which depends
crucially on the choice of version. Hopefully, theoretical and
phenomenological constraints will be brought to bear on this by
future studies. It remains to be seen whether a version of BIMOND
can be found that pass muster given all such requirements. Recent
discussions of matter-of-principle questions, such as the causal
structure of bimetric theories of a different type (where the
interaction term is a function of the metrics themselves, not their
derivatives) can be found, e.g., in \cite{bdg06,bdg07}.
\par
Finally, it has to be realized that however useful such theories may
turn out to be, they must be only effective, approximate theories,
as evinced by the appearance of the a priori unspecified function
$\M$ and the length $\ell$ (or the MOND acceleration $\az$). These
will, hopefully, be calculated from a theory at a deeper stratum.

\section*{Acknowledgements}
I am grateful to Jacob Bekenstein and Luc Blanchet for useful
comments. This research was supported by a center of excellence
grant from the Israel Science Foundation.
\appendix

\clearpage


\begin{thebibliography}{}
\bibitem{milgrom83}M. Milgrom, ApJ, 270, 365 (1983).
\bibitem{bm84}J. Bekenstein and M. Milgrom, ApJ, 286, 7 (1984).
\bibitem{bek04}J.D. Bekenstein, Phys. Rev. D70, 083509 (2004).
\bibitem{sanders97}R.H. Sanders, ApJ, 480, 492 (1997).
\bibitem{bek06}J.D. Bekenstein, Contemp. Phys., 47, 387 (2006).
\bibitem{z06}T.G. Zlosnik, P.G. Ferreira, and G.D. Starkman,
 Phys. Rev. D74, 044037 (2006).
\bibitem{z07}T.G. Zlosnik, P.G. Ferreira, and G.D. Starkman,
 Phys. Rev. D75, 044017 (2007).
\bibitem{skordis09}C. Skordis, Class. Quant. Grav. 26 (14),
 143001 (2009).
\bibitem{blt08}L. Blanchet and A. Le Tiec, Phys. Rev. D78,
024031 (2008).
\bibitem{blt09}L. Blanchet and A. Le Tiec, Phys. Rev. D80,
023524 (2009).
\bibitem{bl07}L. Blanchet, Class. Quant.Grav. 24, 3541 (2007).
\bibitem{milgrom09c}M. Milgrom, MNRAS, in press,
 arXiv:0911.5464 (2009).
\bibitem{rosen74}N. Rosen, Ann. Phys. 84, 455 (1974).
\bibitem{boulanger01}N. Boulanger, T. Damour, L. Gualtieri,
 and M. Henneaux, Nucl. Phys. B, 597, 127 (2001).
\bibitem{dk02}T. Damour and I.I. Kogan, Phys. Rev. D66,
104024 (2002).
\bibitem{bdg06}D. Blas, C. Deffayet, and J. Garriga,
Class. Quant. Grav. 23, 1697 (2006).
\bibitem{bdg07}D. Blas, C. Deffayet, and J. Garriga,
Phys. Rev. D76, 104036 (2007).
\bibitem{banados09}M. Ba\~{n}ados, P.G. Ferreira, and
C. Skordis, Phys. Rev. D79, 063511 (2009).
\bibitem{woodard07} R.P. Woodard, Lect. Notes Phys., 720, 403
(2007).
\bibitem{milgrom09b}M. Milgrom, ApJ, 698, 1630 (2009).
\bibitem{milgrom09a}M. Milgrom, MNRAS, 399, 474 (2009).
\bibitem{sj06}M. Sereno and Ph. Jetzer, MNRAS, 371, 626 (2006).
\bibitem{sanders05}R.H. Sanders, MNRAS, 363, 459 (2005).
\bibitem{bruneton07}J.P. Bruneton and G. Esposito-Farese,
 Phys. Rev. D76, 124012 (2007).
\bibitem{sagi09}E. Sagi, Phys. Rev. D80, 044032 (2009).
\bibitem{z08}T.G. Zlosnik, P.G. Ferreira, and G.D. Starkman,
 Phys. Rev. D77, 084010 (2008).

\end{thebibliography}
\end{document}